\documentstyle[12pt]{article}

\large
\newcommand{\be}{\begin{eqnarray}}
\newcommand{\ee}{\end{eqnarray}}
\newcommand{\bfl}{\begin{flushleft}}
\newcommand{\efl}{\end{flushleft}}
\newcommand\ie {{\it i.e. }}

\newcommand\etal {{\it et.al. }}

\newcommand\noi {\noindent}

\begin{document}

\bibliographystyle{nphys}

\pagestyle{empty}
\vfill
\eject

\vskip 3cm

\vspace* {-25 mm}
\begin{flushright}
USITP-93-6 \\
SUNY-NTG-93-4 \\
March 1993
\end{flushright}
\vskip 1cm
\centerline{\bf String Tension in QED from  Coulomb Gauge Correlators}
\vskip 4cm
\centerline{\bf{T.H. Hansson$^\dagger$ and   I. Zahed$^\star$} }
\vskip 20mm\noi
\centerline{\bf ABSTRACT}
\vskip 5mm

We discuss the new Coulomb gauge method for testing confinement and measuring
the string tension in the context of the Schwinger model and compact QED in 3
dimensions.

\vfil\noi
$^\dagger$
Department of Physics, University of Stockholm \\
Box 6730, S-113 85, Stockholm, Sweden \\
email: hansson@vana.physto.se
\vskip 2mm
\noi$^\star$Nuclear theory group, Department of physics, SUNY at  Stony Brook
\\
Stony Brook, New York, 11794, USA \\
email: zahed@sbnuc1.physics.edu

\vskip 3mm \noindent
${^\dagger}$Supported by the Swedish Natural Science Research Council. \\
${^\star}$Supported in part by the Department of Energy under Grant No.
DE-FG02-88ER40388.
\eject

There are two standard ways to  define confinement and
measure the string tension. The first consists of calculating expectation
values of Wilson loops, and the other of measuring correlation functions
of Polyakov loops (or Wilson lines).
In the confined phase of a pure gauge theory these functions fall
off exponentially with the area enclosed (by the loop in the first case
and between the loops in the second).
In the presence of fermions, neither of the methods provides a test for
confinement.

Recently Marinari \etal proposed to use the correlation
function of time-like gauge fields in Coulomb gauge as an alternative
way to measure the string tension, and they also argued
that their method will be a good test for
confinement even when fermions are present. By a lattice simulation of
pure SU(3) Yang-Mills theory they demonstrated that their method is
efficient for measuring the string tension.

Here we consider this method in the context of abelian models, in particular
the Schwinger model and 3 dimensional compact QED. The first model is
interesting since the fermions completely screens the perturbative
linear potential, and the other since it is confining due to
monopoles.\cite{poly1}. Following Ref. \cite{PARISI2} we
consider the (Euclidian) correlation function between time-like lines
(gaugeon) in Coulomb gauge,
\be
C(T,L) = \langle e^{ie\int_0^T d\tau A_0 (\tau, \vec 0)} \,
  e^{-ie\int_0^T d\tau A_0 (\tau, \vec L)}\rangle \ \ \ \ \ ,
\label{gaugeon}
\ee
where $A_0$ is the time component of the gauge field, and $\vec L =
(L,0,0....)$.
In Coulomb gauge ($\vec\nabla\cdot \vec A=0$) the gauge configurations are
specified up to a time dependent gauge
transformation (assuming that all fields fall off asymptotically). These
transformations leave (\ref{gaugeon}) invariant. Since the theory is abelian,
there is no problem with Gribov copies.\footnote{
For non-Abelian theories
$C(T,L)$ transforms covariantly under the time-dependent residual
gauge transformations, \ie $C(T,L)\rightarrow \Lambda (T)
C(T,L)\Lambda^\dagger(T)$. Note, however, that the  eigenvalues of
$C(T,L)$ are gauge invariant.}
In Coulomb gauge, the longitudinal electric field\footnote{
The calculations are in Euclidian space, so we use the terms electric,
transverse and longitudinal only to keep track of the 0 direction used to
define the Coulomb gauge.}
is given by
$\vec E_L=\vec\nabla A_0$, so (\ref{gaugeon}) can be written in a manifestly
gauge-invariant way as
\be
C(T,L) =  \langle e^{-ie\int_0^Td\tau \int_0^L  dx_1
 E_L^1(\tau , x_1,0,0,..)}\rangle =
e^{-ie\int_0^{T}d\tau \, V_C({\vec x})}
\label{cou}
\ee
where the second equality involving the instantaneous Coulomb potential,
$V_C(\vec x) = 1/\nabla^2$, holds for  non-interacting theories only.
{}From (\ref{cou}) it follows that in the non-interacting case the correlator
$C$
simply measures the Coulomb potential.

In 2 dimensions there is no transverse electric field, thus
$\int_0^T d\tau\int_0^L dx \, E_L(\tau , x) = \oint \, dx^\mu\, A_\mu$,
where the
integral is around the rectangle $[0,T]\times [0,L]$, so $C$ is precisely the
Wilson loop. If there are no charged fields there will be a string tension
$e^2/2$, but in the presence of charges, this will be screened. Taking
two-dimensional QED, \ie the Schwinger model, we can use the standard
procedure to integrate out the
fermions and get a quadratic action for the electric field
\be
L =\frac 12 E^2 -\frac{e^2}{\pi} E\frac 1 \Box E
\label{schwin}
\ee
The expectation value in (\ref{cou}) is now Gaussian and the result is
\be
C(T,L) = e^{-\frac{e^2}{4\pi} \int\int d^2y\, d^2z\, \Box
K_0 (e|y-z|/\sqrt{\pi})}
\label{result}
\ee
For $e=0$ we obtain $C(T,L)=e^{-\frac {e^2} 2 LT}$ (area law), since
$\lim_{e\rightarrow 0} \Box K_0 = 2\pi\delta^2(x-y)$.
For $e\neq 0$ one can  partially integrate (4) to get $C(T,L)=
e^{-e\frac {\sqrt{\pi}} 2  (L+T)}$ (perimeter law). This is to be compared
with  the (connected) correlation function of two Polyakov loops for
$e\neq 0$ which has the
behaviour $\sim e^{-e\frac {\sqrt{\pi}} 2  T }$ (the $L$ dependence disappear
because of the periodicity in the imaginary time direction). We se that the
correlator (\ref{gaugeon}) falls of exponentially in $L$ while the Polyakov
loop correlator is constant. Whether or not this will persist in higher
dimensions so that (\ref{gaugeon}) can be used to establish confinement even
in the presence of light quarks, is an open question.

Polyakov has shown that the presence of monopoles in 3 dimensional compact QED
decorrelates Wilson loops and produces confinement. In this case the correlator
(\ref{gaugeon}) is not the same as the Wilson loop since it is insensitive to
the transverse electric field. In Polyakov's approach, the semi-classical
electric fields  due to monopoles are given by $E^i = \epsilon^{ij}
(2\pi \partial_i/\partial^2)\rho$, where $\rho$ is the monopole density.
Since this field is purely transverse it will, at least to
lowest order, not influence the correlator (\ref{gaugeon}), and we would not
expect an area law. For this abelian model it would be interesting to
make a lattice measurement
not only of (\ref{gaugeon}), but also of the correlator
\be
\tilde C(T,L) = \langle e^{ie\int_0^L dx_1 A_1 (0, x_1,0)} \,
  e^{-ie\int_0^L dx_1 A_1 (T,x_1,0)}\rangle
\ee
which is related to the transverse electric field and thus gauge invariant.

The new method proposed in \cite{PARISI2} provides an
interesting alternative test for confinement and measure of the string
tension. Judging from the Schwinger model we expect
(\ref{gaugeon}) to be sensitive to screening, just like the Wilson loops.
There is, however, the interesting possibility that confinement will be
signalled by an exponential fall-off in (\ref{gaugeon}) even in the presence
of screening. We also propose 3 dimensional compact QED as a good testing
ground for the new
method.

\vglue .5cm
{\bf \noindent  Acknowledgements } \\
\noi
We thank G. Parisi for a useful correspondence.
This work was partially supported by the US DOE grant DE-FG-88ER40388.

\vglue 1cm

%\bibliographystyle{aip}
%\bibliography{vacref}

\begin{thebibliography}{1}

%\bibitem{PARISI1}
%G. Parisi
%\newblock {\it A short Introduction to Numerical Simulations of Lattice gauge
%Theories, in Phenomenes Critiques, Systemes Aleatoires, Theories de Jauge,}
%eds. K. Osterwalder and R. Stora (North-Holland) 1986.


\bibitem{PARISI2}
E. Marinari, M.L. Paciello, G. Parisi and B. Taglienti,
\newblock Phys. Lett. {\bf 298B}, 400 (1993).

\bibitem{poly1}
A. M. Polyakov,
\newblock Nucl. Phys. {\bf B120}, 429 (1977).

\end{thebibliography}

\end{document}